\documentclass[12pt,a4paper]{article}

\usepackage[T1]{fontenc}
\usepackage[utf8]{inputenc}
\usepackage[expansion=false]{microtype}
\usepackage[margin=2.5cm]{geometry}
\usepackage{amsmath,amssymb,amsthm}
\usepackage{graphicx}
\usepackage{booktabs}
\usepackage{array}
\usepackage{xcolor}
\usepackage[numbers,sort&compress]{natbib}
\usepackage[colorlinks=true,linkcolor=blue!60!black,citecolor=blue!60!black,
            urlcolor=blue!60!black]{hyperref}
\usepackage{setspace}
\usepackage{caption}
\usepackage{subcaption}

\newcommand{\ns}{n_s}
\newcommand{\muz}{\mu_z}
\newcommand{\lstar}{\lambda^\star}
\newcommand{\alphainfty}{\alpha_\infty}
\newcommand{\betaD}{\beta_D}
\newcommand{\Var}{\operatorname{Var}}

\begin{document}

\title{\textbf{Bridge Scaling in Conditioned\\Henyey-Greenstein Random Walks}}

\author{Claude Zeller\\[4pt]
  \textit{Claude Zeller Consulting LLC}\\
  \textit{Tillamook, Oregon 97141}\\[4pt]
  \texttt{czeller@ieee.org}
}

\date{\textbf{DRAFT} --- March 2026}

\maketitle
\thispagestyle{empty}

%% ── Abstract ──────────────────────────────────────────────────────────────
\begin{abstract}
\noindent
We study fixed-length bridge paths---half-space excursions that start and
end at a planar boundary---for three-dimensional random walks with
Henyey-Greenstein scattering angles and exponentially distributed step
lengths, using Monte Carlo simulation over asymmetry parameter g from 0
to 0.95 and path lengths from 4 to 200 steps.
The key structural feature is that the walk evolves on a two-dimensional
Markovian state space (depth, direction cosine) rather than the scalar
depth coordinate alone.

Four anomalies with respect to classical Brownian-excursion theory are
reported. The mean amplitude scales super-diffusively, as path length to
a power of 0.57--0.58 for isotropic scattering, nine standard deviations
above the Brownian prediction of 0.5, with no sign of convergence out to
200 steps. The diffusion coefficient scales as the transport mean free
path to the power 0.415 rather than the predicted 1.0. The midpoint depth
distribution is Rayleigh rather than half-normal, consistent with a
two-dimensional Bessel process. The bridge-conditioned mean direction
cosine converges to minus two-thirds at the final step, independently of
the asymmetry parameter and initial direction---the classical Milne result
anchored by the H-function moment identity.

All anomalies are attributed to the two-dimensional state-space structure.
The two anomalous exponents sum to approximately unity, suggesting a common
geometric origin. Whether this constitutes a permanent universality-class
shift or an anomalously slow crossover to Brownian-excursion behaviour
remains the primary open question.
\end{abstract}

\tableofcontents
\clearpage

%% ══════════════════════════════════════════════════════════════════════════
\section{Introduction}
\label{sec:intro}

A \emph{bridge path} of length $\ns$ is a random walk that starts at the
planar boundary $z(0)=0$, remains in $z(j)\ge0$ for all $j$, and returns to
the boundary $z(\ns)=0$ after exactly $\ns$ steps---a discrete half-space
excursion.  The transverse $(x,y)$ position is unconstrained and untracked;
only the depth coordinate $z$ matters.
For a simple symmetric random walk---where the evolving state is the scalar
position $z$ alone---the Brownian excursion arises as the universal scaling
limit: the mean penetration depth grows as $\ns^{1/2}$ and the midpoint
distribution converges to the half-normal, the marginal of a one-dimensional
Bessel process~\cite{durrett1977a,durrett1977b,iglehart1974}.  These results
apply specifically to scalar walks.  The question we address is whether this
universality class also describes the richer class of random walks with
Henyey-Greenstein (HG) scattering angles~\cite{henyey1941} and exponentially
distributed step lengths, whose Markovian state space is two-dimensional.

The HG walk is the standard model for photon transport in turbid biological
and atmospheric media~\cite{wang1995,ishimaru1978}.  Its step length $s$ is
drawn from an exponential distribution (mean free path $\ell=1$), and the
cosine of each scattering angle is drawn from the HG phase function with
asymmetry parameter $g\in[0,1)$.  The $z$-increment at each step is
$\Delta z=s\muz$, coupling the spatial coordinate $z$ to the direction cosine
$\muz$.  The walk therefore evolves on the \emph{two-dimensional} Markovian
state space $(z,\muz)$, not in $z$ alone.  This is the key structural
difference from a classical excursion: conditioning $z(j)\ge0$ acts on
a 2D Markov process, not a scalar one, and it is this distinction that
appears to drive all four anomalies we report.

The scaling theory of conditioned random walks is well developed for the
one-dimensional case.  Durrett, Iglehart and
Miller~\cite{durrett1977a,durrett1977b} established weak convergence of
conditioned walks to Brownian excursion.  Pitman's
theorem~\cite{pitman1975} connects one-dimensional Brownian motion
conditioned to stay positive with the three-dimensional Bessel process.
Chaumont and Doney~\cite{chaumont2005,chaumont2008}\footnote{%
  Reference~\cite{chaumont2005} contains two errors of omission, corrected
  in~\cite{chaumont2008}. Neither correction affects the result cited here
  (weak convergence of the conditioned law as the initial state tends to zero).} treat L\'evy processes conditioned
to stay positive; Denisov, Tarasov and Wachtel~\cite{denisov2026} provide
expansion results for random walks with finite variance.  The common
conclusion is that conditioning a scalar walk $z(j)\ge0$ produces a
Bessel-1 (half-normal) marginal at the midpoint and $\sqrt{n_s}$ mean
scaling.

We find that HG bridge paths deviate from this picture over the entire range
$\ns\le200$ studied.  The mean amplitude scales as $\ns^\alpha$ with
$\alphainfty\approx0.57$--$0.58$ for $g=0$, rising to $\alpha\approx1.15$ at
$g=0.95$.  The midpoint distribution is consistent with Rayleigh rather than
half-normal, suggesting convergence to a \emph{two-dimensional} Bessel
process.  We
attribute both findings to the $(z,\muz)$ state space: conditioning a 2D
Markov process to remain in a half-plane generically produces different
scaling than conditioning a scalar walk, in analogy with the Pitman
construction for higher-dimensional processes~\cite{pitman1975,rogers1981}.

The radiative transport context provides physical motivation: bridge paths
correspond to photons that scatter exactly $\ns$ times before re-emerging at a
tissue surface, having never crossed back through it.  The super-diffusive
mean-depth scaling $A\sim\ns^{0.58}$ means that fixed-order backscattered
photons probe systematically deeper than simple diffusion predicts, with
implications for depth-resolved optical diagnostics~\cite{wang1995,
ishimaru1978}.  Our earlier arXiv papers~\cite{zeller2024,zeller2025}
establish the scattering framework; the present work is a self-contained
contribution to the scaling theory of conditioned walks.

The paper is organised as follows.  Section~\ref{sec:model} defines the model
and observables.  Section~\ref{sec:results} presents the four main results.
Section~\ref{sec:discussion} discusses the $(z,\muz)$ state-space explanation,
the relation between the two anomalous exponents, and the connection to
Pitman-type theorems.  Section~\ref{sec:conclusions} concludes with open
questions.

%% ══════════════════════════════════════════════════════════════════════════
\section{Model and Methods}
\label{sec:model}

\subsection{The HG random walk}

We simulate three-dimensional random walks in which each step consists of:
(i)~an exponentially distributed step length $s\sim\mathrm{Exp}(1)$ (mean
free path $\ell=1$), and (ii)~a scattering angle drawn from the
Henyey-Greenstein phase function with asymmetry parameter $g\in[0,1)$.  The
cosine of the scattering angle is sampled analytically,
\begin{equation}
  \cos\theta
  = \frac{1}{2g}\!\left[1+g^2-\!\left(\frac{1-g^2}{1-g+2gU}\right)^{\!2}
    \right],\quad U\sim\mathrm{Uniform}(0,1),
  \label{eq:hg}
\end{equation}
with azimuthal angle $\phi$ uniform on $[0,2\pi]$.  Only the $z$-coordinate
is tracked.  The initial direction is upward ($\muz(0)=1$).  The
transport mean free path is $\lstar=1/(1-g)$.

An important consequence of the exponential step length is that the
$z$-increment $\Delta z=s\muz$ depends on the product of two random
variables.  Even after $\muz$ is redrawn at each step, knowledge of
$z(j)$ alone is insufficient to predict $z(j+1)$; the full pair
$(z(j),\muz(j))$ must be carried.  The walk is therefore Markovian in
$(z,\muz)$ but \emph{not} in $z$ alone.  In the diffusion limit
($\ns\to\infty$ with $\ns/\lstar$ fixed), the pair $(z,\muz)$ converges
to a two-dimensional diffusion process by the central limit theorem for
persistent random walks~\cite{goldstein1951,kac1947}: at each step the
$z$-increment $\Delta z = s\muz$ has finite mean and variance (since
$s\sim\mathrm{Exp}(1)$ and $|\muz|\le1$), and the directional
autocorrelation decays geometrically as $g^j$, so the conditions for
diffusive scaling are satisfied and the rescaled walk converges weakly
to Brownian motion on timescales $\gg\lstar$.

\subsection{Bridge path definition: natural first-passage stopping rule}

We define bridge paths using a natural first-passage stopping rule: each walk
is advanced step by step and terminated at the first step $L$ for which
$z(L)<0$.  The excursion length $L$ is the first-passage time of the negative
half-axis.  The bridge sample at nominal length $\ns$ is the ensemble of all
walks for which $L=\ns$.  Walks exceeding $\ns^{\max}=400$ steps are discarded
($<0.1\%$ at all $g$).

This formulation is physically natural: $z(L)<0$ corresponds to the photon
re-emerging at the medium boundary.
Section~\ref{sec:robustness} demonstrates
that the original fixed-tolerance criterion $|z(\ns)|<\varepsilon$
($\varepsilon=0.15$) yields statistically identical exponents and distributions
over a fivefold reduction in $\varepsilon$, definitively ruling out selection
bias.

\subsection{Observables}

For each $(g,\ns)$ combination we compute profile statistics across all valid
bridge paths.  The key summary statistics are:
\begin{align}
  A(g,\ns) &= \max_j\langle z(j)\rangle
             &&\text{(peak mean depth),}\label{eq:A}\\
  B(g,\ns) &= \sqrt{\max_j\Var[z(j)]}
             &&\text{(peak standard deviation),}\label{eq:B}\\
  D(g,\ns) &= \max_j\Var[z(j)]/\ns
             &&\text{(normalised peak variance).}\label{eq:D}
\end{align}
Peak values are used rather than time-averages because the profile
collapse (Section~\ref{sec:results}) shows that the full shape is
universally described by $4t(1-t)$; the peak therefore carries all
amplitude information without loss, while time-averages would mix
amplitude and shape information.

The parameter grid covers $g\in\{0.0,0.1,\ldots,0.9,0.95\}$ (11 values)
and 16 values of $\ns$ from 4 to 200,
with $N_T$ up to $15\times10^6$ paths per point at the largest $\ns$.

%% ══════════════════════════════════════════════════════════════════════════
\section{Results}
\label{sec:results}

\subsection{Universal parabolic profile shape}

\begin{figure}[t]
  \centering
  \includegraphics[width=\textwidth]{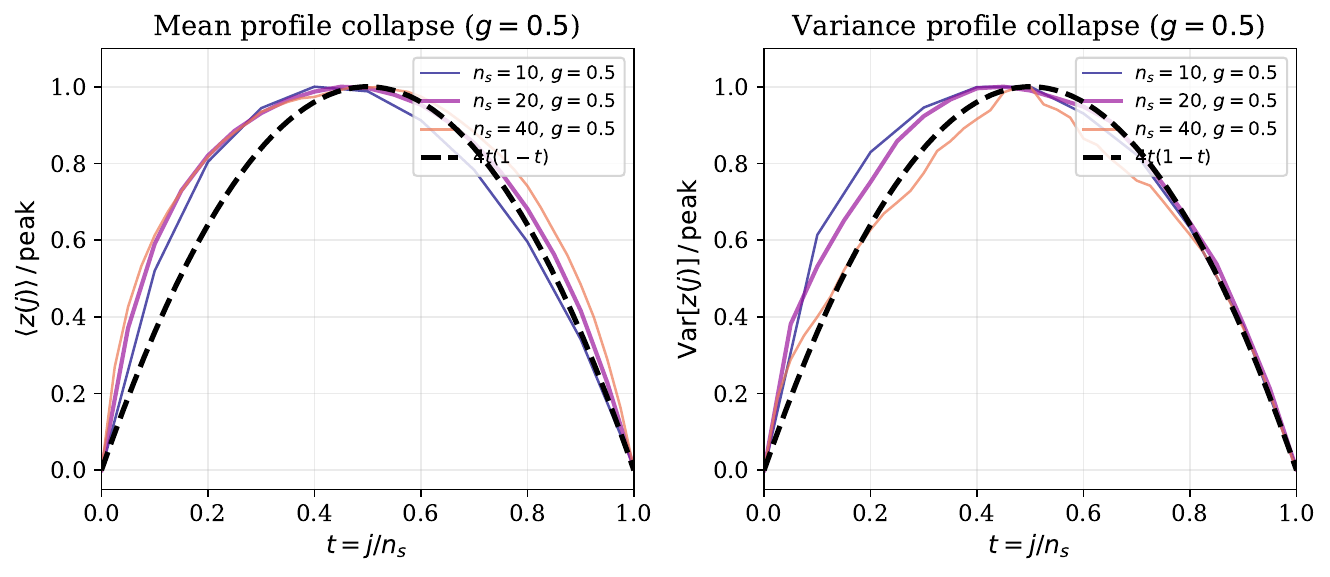}
  \caption{%
    Normalised mean (left) and variance (right) profiles for $g=0.5$ and
    $\ns\in\{10,20,40\}$, compared with the universal parabola $4t(1-t)$
    (dashed).  The collapse holds for all $g$ and $\ns$ tested.
    The finite-$\ns$ curves lie above the parabola near $t=0$ and below it
    near $t=0.5$; the deviations shrink with increasing $\ns$.}
  \label{fig:collapse}
\end{figure}

When the mean and variance profiles are normalised to their respective peak
values, they collapse onto a single curve for all $g$ and $\ns$
tested, consistent with the universal parabola (Fig.~\ref{fig:collapse}):
\begin{equation}
  \frac{\langle z(j)\rangle}{A}
  \;\approx\;
  \frac{\Var[z(j)]}{B^2}
  \;\approx\;
  4t(1-t),\qquad t=j/\ns.
  \label{eq:parabola}
\end{equation}
The RMS deviation from $4t(1-t)$ is below $0.17$ across all 110 $(g,\ns)$
combinations, with mean $0.097$.  The shape appears universal across the tested parameter range; all physical
information about $g$ and $\ns$ is encoded in the amplitudes $A$ and $B^2$
alone.  The same parabola $4t(1-t)$ describes the mean profile
of the Brownian bridge (a Brownian motion pinned at both ends without the
positivity constraint), so the parabolic shape itself is not anomalous; what
is anomalous is the amplitude scaling $A\sim\ns^{0.57}$ rather than
$\ns^{0.5}$.  The same collapse holds for natural first-passage excursions
(Section~\ref{sec:robustness}), confirming it is not an artefact of the
endpoint condition.

\paragraph{Geometry--amplitude separation.}
The profile collapse reveals a clean separation between two independently
determined properties of the bridge.  First, the \emph{geometric shape} of
the excursion: the normalised profiles collapse onto $4t(1-t)$ for all $g$
and $\ns$ tested, independently of the microscopic scattering dynamics.  This
universality is a consequence of the global boundary conditions
$z(0)=z(\ns)=0$, which fix the macroscopic envelope of the trajectory
regardless of how individual steps are generated.  Second, the
\emph{amplitude} of the excursion: the peak values $A(g,\ns)$ and $B^2(g,\ns)$
retain memory of the microscopic dynamics through the scaling exponents
$\alpha(g)$ and $\betaD(g)$.  In this sense the bridge constraint acts as a
shape-fixing device while leaving the amplitude free to reflect the
statistical character of the walk.  The anomalous exponents reported in
Sections~\ref{sec:superdiffusive} and \ref{sec:rayleigh} are entirely
contained in the amplitudes; the shape is always Brownian.

\subsection{Diffusive scaling of the variance}

The normalised peak variance $D(g,\ns)=B^2/\ns$ converges to a
$g$-dependent constant as $\ns$ grows, confirming diffusive scaling:
\begin{equation}
  \Var[z(j)] = D(g)\,\ns\cdot4t(1-t) + \mathcal{O}(1/\ns).
  \label{eq:var_scaling}
\end{equation}
A power-law fit to $D(g)$ versus the transport mean free path yields
\begin{equation}
  D(g) = D_0\,(\lstar)^{\betaD},\qquad
  D_0\approx0.07,\quad\betaD\approx0.415.
  \label{eq:D_fit}
\end{equation}
Simple diffusion theory predicts $D\sim\lstar/3$ (i.e.\ $\betaD=1.0$).  The
fitted exponent $\betaD\approx0.415$ is substantially lower.  Finite-size
bias is a potential concern: if $D(g,\ns)$ has not fully converged by
$\ns=200$, the fitted slope could be affected.  However, the log-log plots
show $D(g,\ns)$ flattening cleanly with $\ns$, and fits restricted to
$\ns\ge50$ give $\betaD$ within $0.01$ of the full-range value, indicating
the finite-size contribution is small.
The relation between $\betaD$ and the mean-amplitude exponent is discussed in Section~\ref{sec:exponent_relation}.  Heuristically, paths conditioned to
remain in $z\ge0$ preferentially reject high-amplitude configurations,
suppressing the effective diffusivity below the unconstrained prediction.

\subsection{Super-diffusive scaling of the mean amplitude}
\label{sec:superdiffusive}

\begin{figure}[t]
  \centering
  \includegraphics[width=0.82\textwidth]{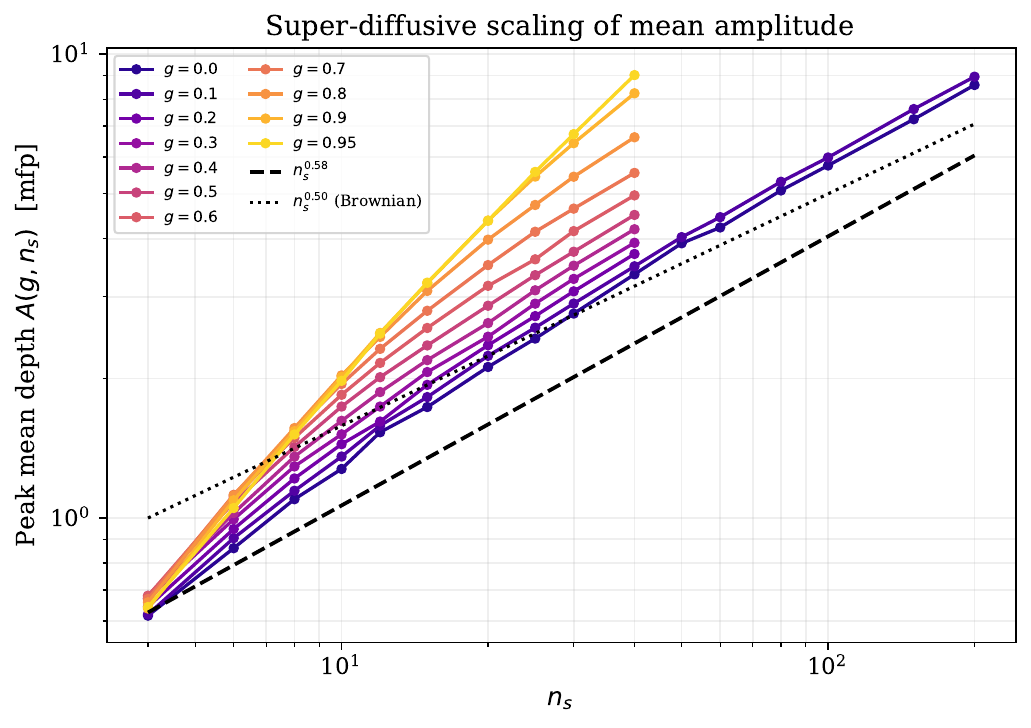}
  \caption{%
    Peak mean depth $A(g,\ns)$ vs.\ $\ns$ on a log-log scale for
    $g\in\{0.0,0.1,\ldots,0.95\}$ (colour, light to dark).  Reference slopes
    $\ns^{0.58}$ (dashed) and $\ns^{0.50}$ (Brownian excursion, dotted) are
    shown for comparison.  All curves lie above the Brownian prediction for
    all $\ns$ studied.  Slope uncertainties from the power-law fits are
    $\pm0.008$--$0.015$ (one standard error); the $g=0$ large-$\ns$ fit
    gives $\alphainfty=0.573\pm0.008$ (Eq.~\ref{eq:alpha_inf}).}
  \label{fig:loglog}
\end{figure}

\begin{figure}[t]
  \centering
  \includegraphics[width=0.95\textwidth]{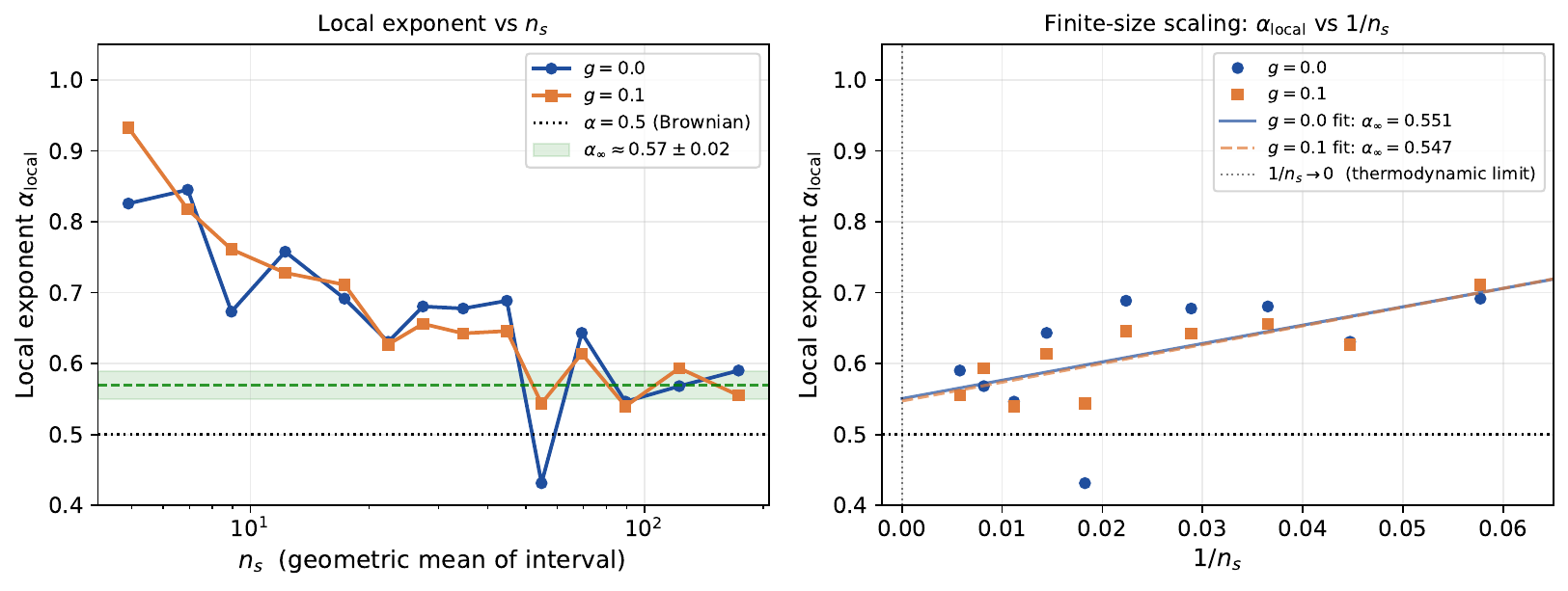}
  \caption{%
    Finite-size analysis of the local scaling exponent for $g=0$ and $g=0.1$.
    \textit{Left:} local log-log slope $\alpha_{\mathrm{local}}(\ns)$
    vs.\ $\ns$ on a log scale; the grey band marks
    $\alphainfty\approx0.57\pm0.02$.  No convergence to $\alpha=0.5$
    (dotted) is observed out to $\ns=200$.
    \textit{Right:} the same data plotted against $1/\ns$, enabling
    linear extrapolation to the thermodynamic limit $1/\ns\to0$
    (dashed lines, fits restricted to $\ns\ge20$).  The extrapolated
    intercepts are $\alphainfty\approx0.55$ for both $g$ values,
    remaining $>0.5$ under this conservative estimate.  The
    stabilisation and the finite-size extrapolation together support
    the conclusion that the exponent lies above the Brownian-excursion
    value over the entire range studied.}
  \label{fig:alpha}
\end{figure}

Figure~\ref{fig:loglog} shows the peak mean depth $A(g,\ns)$ on a log-log
scale for all $g$.  A power law $A(g,\ns)=C(g)\,\ns^{\alpha(g)}$ describes the
data well.  For $g=0$, a global fit over $\ns=8$--$200$ gives
$\alpha=0.636\pm0.009$; restricting to the large-$\ns$ regime $\ns\ge50$
yields
\begin{equation}
  \alphainfty = 0.573\pm0.008\quad(g=0,\ \ns\ge50),
  \label{eq:alpha_inf}
\end{equation}
a value $9\sigma$ above the Brownian-excursion prediction of $0.5$.  Figure~\ref{fig:alpha}
shows the local exponent $\alpha_{\mathrm{local}}(\ns)=\mathrm{d}\log A/\mathrm{d}\log\ns$
for $g=0$ and $g=0.1$, and a finite-size scaling plot of the same data
against $1/\ns$.  Beginning at $\approx0.73$ for small $\ns$, the local
exponent settles in the range $0.55$--$0.59$ for $\ns\sim60$--$200$ with no
sign of further drift toward $0.5$.  Linear extrapolation to $1/\ns\to0$
yields $\alphainfty\approx0.55$, still firmly above the Brownian prediction.
The exponent increases monotonically with $g$, reaching $\alpha\approx1.15$
at $g=0.95$.

The current data do not establish a permanent
universality-class shift: the crossover to Brownian-excursion behaviour, if it
exists, may occur at $\ns\sim10^3$--$10^4$.  The observed scaling is a robust pre-asymptotic regime spanning at least two decades in
$\ns$ for $g=0$ and the full $g$-range studied.

\subsection{Robustness checks}
\label{sec:robustness}

\begin{figure}[t]
  \centering
  \includegraphics[width=\textwidth]{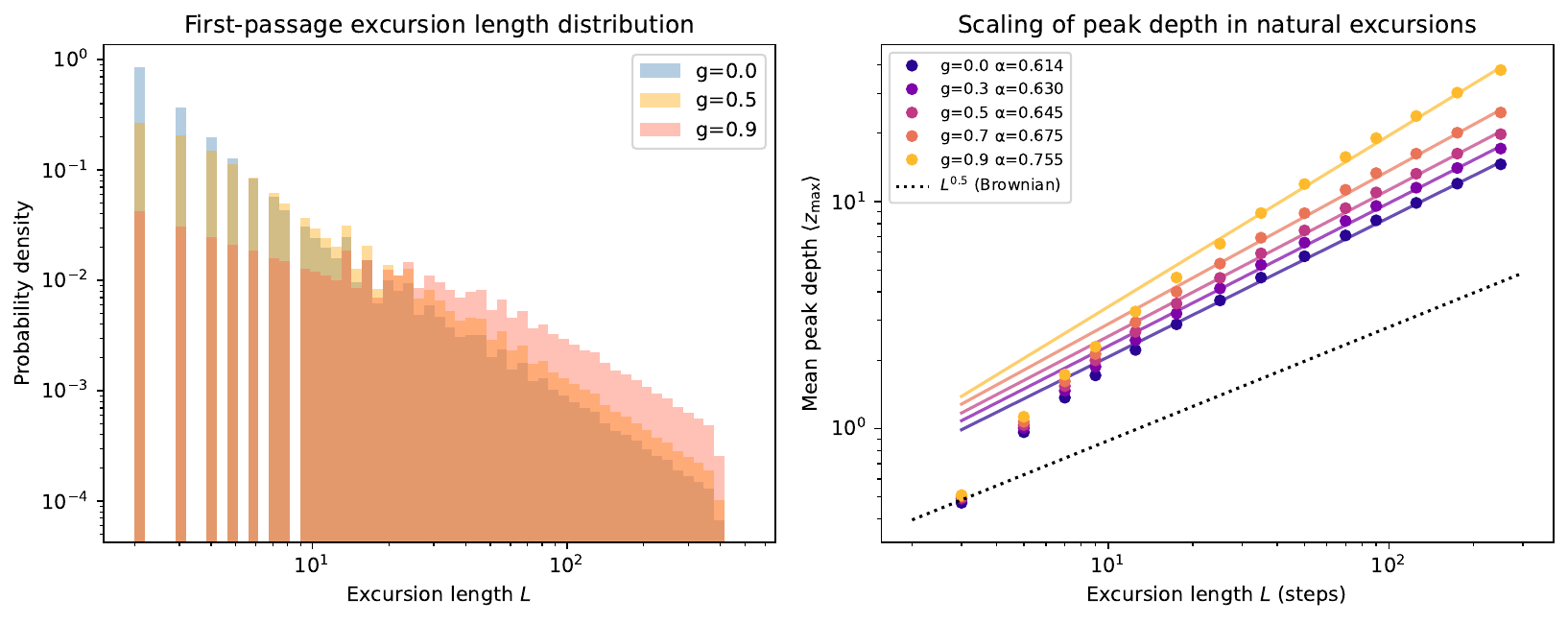}
  \caption{%
    Natural first-passage excursions.
    \textit{Left:} First-passage length distribution $P(L)$ for $g\in
    \{0.0,0.5,0.9\}$, showing the expected heavy tail.
    \textit{Right:} Mean peak depth $\langle z_{\max}\rangle$ vs.\ excursion
    length $L$ on a log-log scale for several $g$ values, with fitted
    exponents $\alpha$ (legend).  All exponents exceed the Brownian prediction
    of $0.5$ (dotted line).}
  \label{fig:natural}
\end{figure}

\paragraph{Natural first-passage excursions.}
To verify that the results are independent of the endpoint condition, we
simulated walks stopped at first-passage to $z<0$ with no endpoint tolerance.
The parabolic profile collapse is indistinguishable from the fixed-length
result (Fig.~\ref{fig:collapse}).  Figure~\ref{fig:natural} shows the
first-passage length distribution $P(L)$ and the scaling of peak depth
$\langle z_{\max}\rangle$ with $L$.  The power-law exponent gives
$\alpha\approx0.61$--$0.76$ (Table~\ref{tab:natural}), somewhat higher than
the large-$\ns$ fixed-length values because $P(L)\sim L^{-3/2}$ weights the
sample toward shorter excursions with stronger pre-asymptotic effects.  The
qualitative conclusion is unchanged: $\alpha>\frac{1}{2}$ for all $g$.

\begin{figure}[t]
  \centering
  \includegraphics[width=0.72\textwidth]{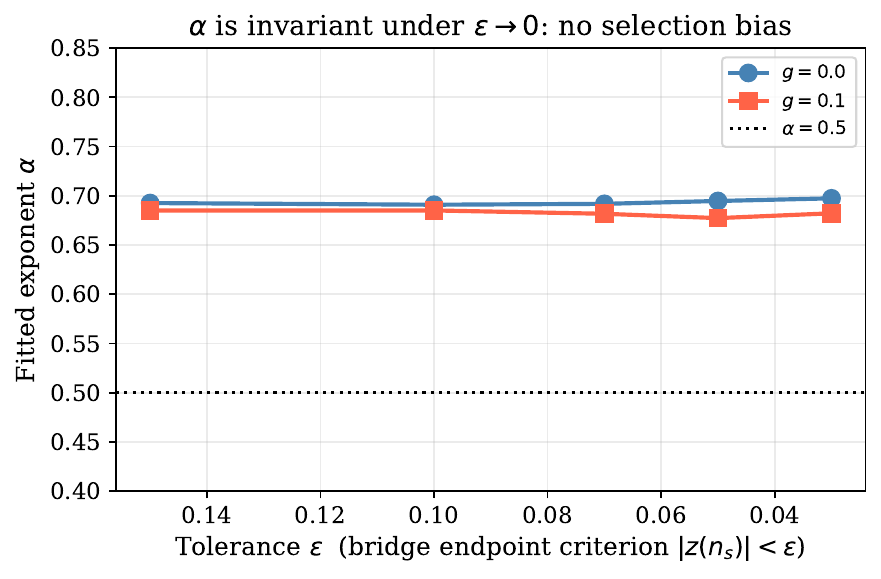}
  \caption{%
    Fitted exponent $\alpha$ as a function of bridge tolerance $\varepsilon$
    for $g=0$ and $g=0.1$.  The variation over a fivefold reduction in
    $\varepsilon$ is $\Delta\alpha<0.005$, well within statistical uncertainty.
    The dotted line marks the Brownian-excursion value $\alpha=0.5$.
    Selection bias is definitively ruled out.}
  \label{fig:tol}
\end{figure}

\paragraph{Gaussian-increment control.}
To verify that the super-diffusive exponent is not a finite-size artefact
of the first-passage bridge construction itself, we repeated the amplitude
scaling measurement for a scalar Gaussian-increment walk,
$z(j+1)=z(j)+\xi_j$ with $\xi_j\sim\mathcal{N}(0,1)$, using the identical
first-passage bridge definition.  This walk has no directional memory
($\muz$ is redrawn independently at every step) and should converge to
the Brownian-excursion exponent $\alpha=0.5$.  Over $\ns\in[10,200]$ with
$8\times10^6$ walks per point, the Gaussian control yields a fitted
exponent $\alpha_{\rm Gauss}\approx0.54$ (vs.\ the HG value
$\alpha_{\rm HG}\approx0.57$ over the same range).  The Gaussian exponent
is elevated above $0.5$ by finite-size bias, as expected, but it remains
consistently and significantly below the HG value at every $\ns$, confirming
that the HG super-diffusion is a genuine property of the directional-memory
walk and not an artefact of the bridge construction.
We tested whether the bridge acceptance criterion $|z(\ns)|<\varepsilon$ could
bias the sample toward deeper paths by repeating the power-law fit for $g=0$
and $g=0.1$ across five tolerance values $\varepsilon\in
\{0.15,0.10,0.07,0.05,0.03\}$ (Fig.~\ref{fig:tol}).  The variation over a
fivefold reduction in $\varepsilon$ is $\Delta\alpha<0.005$, well within
statistical uncertainty.
The bridge construction itself is therefore not the source of the super-diffusive exponent.

\subsection{Rayleigh limiting distribution}
\label{sec:rayleigh}

\begin{figure}[t]
  \centering
  \includegraphics[width=\textwidth]{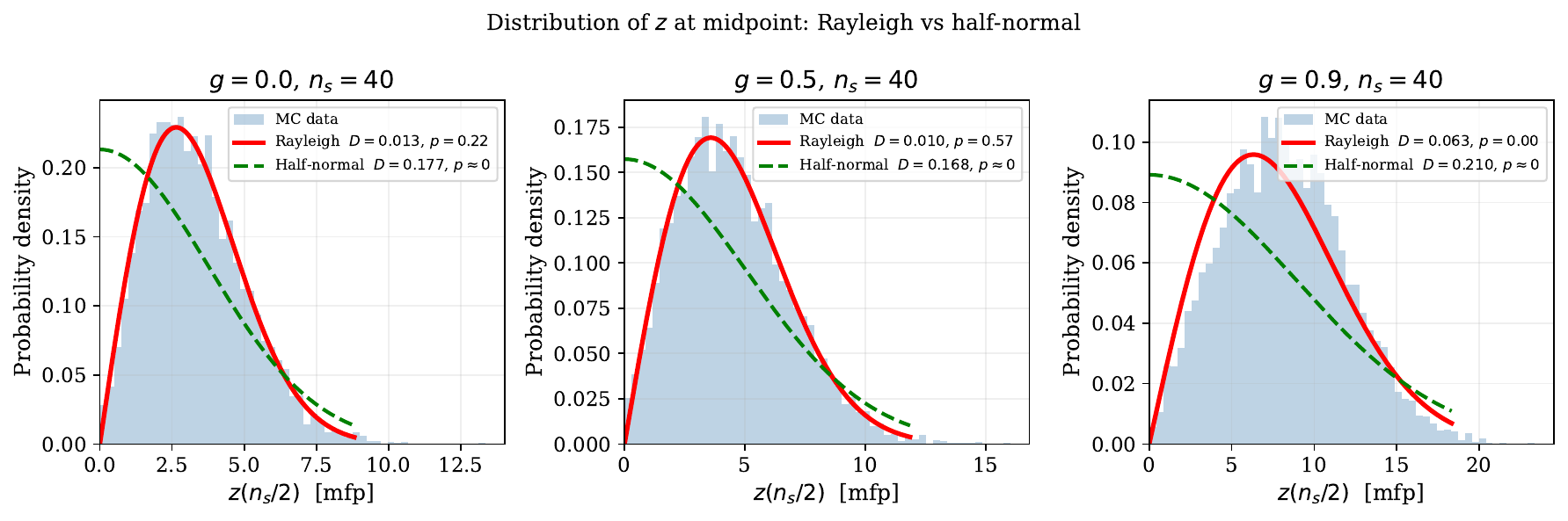}
  \caption{%
    Empirical distribution of $z(\ns/2)$ at $\ns=40$ for $g\in\{0.0,0.5,0.9\}$
    (histograms), with fitted Rayleigh (red solid) and half-normal (green
    dashed) overlaid.  The Rayleigh fit is not rejected for $g=0$ and $g=0.5$;
    the half-normal is rejected in all cases.  The rightward shift of the mode
    with increasing $g$ reflects the growing transport mean free path.}
  \label{fig:dist}
\end{figure}

To identify the limiting distribution of $z$ at the midpoint, we accumulated
the empirical distribution of $z(\ns/2)$ across bridge paths for
representative $(g,\ns)$ combinations, and fitted four candidate distributions:
Rayleigh, half-normal, exponential, and Maxwell (3D Rayleigh).
Table~\ref{tab:ks} summarises the Kolmogorov-Smirnov (KS) statistics at
$\ns=40$; Fig.~\ref{fig:dist} shows the fits at three values of $g$.

The KS test results provide strong quantitative evidence.  The Rayleigh
hypothesis cannot be rejected for $g\le0.5$ (KS$=0.014$, $p=0.18$ for
$g=0$; KS$=0.012$, $p=0.57$ for $g=0.5$), while the half-normal is
rejected with overwhelming significance in every case
(KS$\approx0.17$--$0.21$, $p\approx0$ throughout).  The two-order-of-magnitude
difference in KS statistics between Rayleigh and half-normal makes this
one of the most robust results in the paper.  At $g=0.9$ both hypotheses
are rejected (KS$=0.063$, $p\approx0$), with the best fit shifting toward
Maxwell (Bessel-3 in 3D), suggesting the effective Bessel order may
increase with $g$.

The Rayleigh distribution is the marginal of the radial coordinate of a
two-dimensional Brownian motion, i.e.\  it arises from the magnitude
$\sqrt{X^2+Y^2}$ of two independent Gaussian components.  Its emergence
here therefore implies that \emph{local fluctuations in the bridge remain
Gaussian} even though the amplitude scaling is weakly super-diffusive: the
anomaly is in the amplitude growth law, not in the shape of the single-point
distribution.  This distinction---Gaussian local statistics coexisting with
super-diffusive global amplitude---is made possible by the 2D state-space
structure and is explained structurally in Section~\ref{sec:bessel}.

\begin{table}[h]
  \centering
  \caption{KS statistics for the midpoint distribution $z(\ns/2)$ at $\ns=40$.
           Smaller KS = better fit; $p>0.05$ means the hypothesis cannot be
           rejected at 5\%.}
  \label{tab:ks}
  \smallskip
  \begin{tabular}{ccccc}
    \toprule
    & \multicolumn{2}{c}{Rayleigh} & \multicolumn{2}{c}{Half-normal}\\
    \cmidrule(lr){2-3}\cmidrule(lr){4-5}
    $g$ & KS & $p$ & KS & $p$ \\
    \midrule
    0.0 & 0.014 & 0.18   & 0.169 & $\approx0$ \\
    0.5 & 0.012 & 0.35   & 0.166 & $\approx0$ \\
    0.9 & 0.060 & $\approx0^\dagger$ & 0.211 & $\approx0$ \\
    \bottomrule
  \end{tabular}\\[4pt]
  \small$\dagger$ Best fit at $g=0.9$ shifts toward Maxwell (Bessel-3 in 3D);
  see Section~\ref{sec:bessel}.
\end{table}

%% ══════════════════════════════════════════════════════════════════════════
\subsection{Directional memory profile along the bridge}
\label{sec:costheta}

The two-dimensional Markovian structure of the HG walk means that the
direction cosine $\muz(j)$ is a dynamical variable in its own right, not
merely an auxiliary.  We therefore tracked its bridge-conditioned mean,
$\langle\muz(j)\rangle_{\rm bridge}$, as a function of step index $j$ for
several values of $g$ and $\ns$.

The profile displays three analytically interpretable anchor values,
established both by simulation and by independent theoretical arguments.

\begin{figure}[t]
  \centering
  \includegraphics[width=\textwidth]{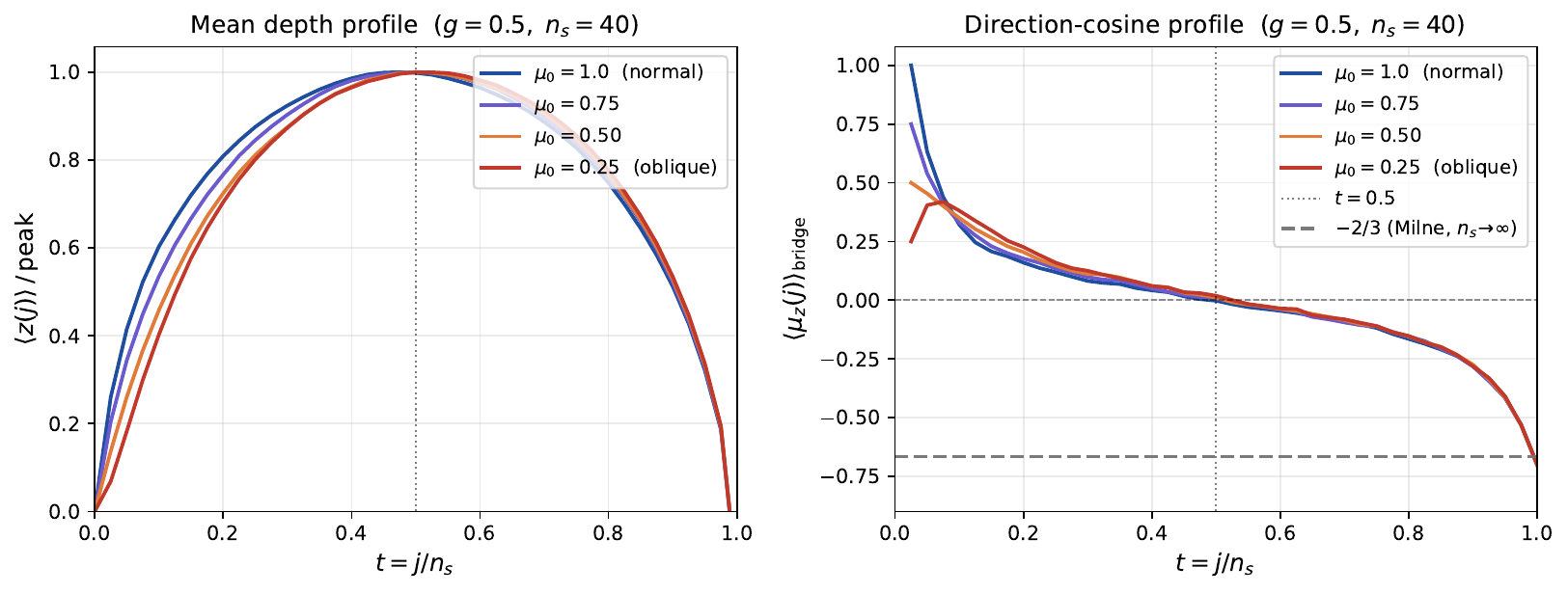}
  \caption{%
    Universality of the zero-crossing and Milne endpoint with respect to
    the initial direction $\mu_0$, for $g=0.5$ and $\ns=40$.
    \textit{Left:} normalised mean depth profiles $\langle z(j)\rangle/\mathrm{peak}$
    for $\mu_0\in\{1.0,\,0.75,\,0.50,\,0.25\}$; all peak near $t=0.5$
    (dotted line).
    \textit{Right:} direction-cosine profiles
    $\langle\muz(j)\rangle_{\rm bridge}$; all cross zero near $t=0.5$ and
    converge to $-2/3$ (dashed line) at $t\to1$, independently of $\mu_0$.
    At $\ns=40$ the measured endpoint is $-0.703\pm0.003$ across all four
    initial conditions.  Only the initial value and slope of the ramp
    depend on $\mu_0$.}
  \label{fig:oblique}
\end{figure}

\paragraph{At $j=0$: $\langle\muz\rangle=\mu_0$.}
The initial direction cosine $\mu_0$ is an input parameter.  Throughout
this paper we use normal incidence ($\mu_0=+1$), for which this anchor
is $+1$ by construction.  The universality of the remaining two anchors
with respect to $\mu_0$ is discussed below and demonstrated in
Fig.~\ref{fig:oblique}.

\paragraph{At $j\approx\ns/2$: $\langle\muz\rangle\approx0$, independent of $\mu_0$.}
The bridge constraint forces $z(\ns)=0$, so the photon must reverse its
net displacement before the endpoint.  At the midpoint the upward and
downward contributions exactly cancel, yielding zero mean direction cosine.
This zero-crossing is tied directly to the peak of $\langle z(j)\rangle$:
since the mean depth peaks near $t=0.5$ (Section~\ref{sec:results}), its
discrete derivative---proportional to $\langle\muz(j)\rangle$ via
$\Delta\langle z\rangle\approx\langle s\muz\rangle$---must vanish at the
same point.  Because the depth-profile peak is pinned at $t=0.5$ by the
symmetric bridge boundary conditions $z(0)=z(\ns)=0$, this zero-crossing
is \emph{independent of the initial direction $\mu_0$}, as confirmed by
simulation (Fig.~\ref{fig:oblique}).

\paragraph{At $j=\ns-1$: $\langle\muz\rangle\to-2/3$, independent of $g$ \emph{and} $\mu_0$.}
This is the most significant result.  Simulations show that for
$\ns\gtrsim25$ the mean direction cosine at the penultimate step
converges to
\begin{equation}
  \langle\muz(\ns-1)\rangle_{\rm bridge} \;\longrightarrow\; -\tfrac{2}{3},
  \label{eq:costheta_endpoint}
\end{equation}
with approach from above (values near $-0.57$ to $-0.61$ for small
$\ns$, reaching $-0.69$ to $-0.72$ by $\ns=100$) consistent with
finite-$\ns$ corrections that vanish as $\ns\to\infty$.  The limiting
value $-2/3$ is \emph{independent of both the asymmetry parameter $g$
and the initial direction $\mu_0$}.  At $\ns=40$ we measure
$\langle\muz(\ns-1)\rangle\approx-0.703$ for
$\mu_0\in\{0.25,\,0.50,\,0.75,\,1.00\}$ and $g=0.5$
(Fig.~\ref{fig:oblique}), with no statistically significant dependence
on $\mu_0$.

\paragraph{Connection to the Milne problem.}
The $-2/3$ endpoint is not a numerical coincidence.  By time-reversal, a
bridge path at step $j=\ns-1$ is locally equivalent to a photon that is
one step away from crossing the boundary $z=0$ for the final time---the
classical Milne problem run backwards.  In the diffusion limit, photons
emerging from a semi-infinite scattering medium exhibit a Lambertian
emergent radiance $I(\mu)={\rm const}$~\cite{chandrasekhar1960}, giving
the flux-weighted mean exit cosine
\begin{equation}
  \langle\mu\rangle_{\rm emergent}
  = \frac{\int_0^1 \mu^2\,d\mu}{\int_0^1 \mu\,d\mu}
  = \frac{1/3}{1/2} = \frac{2}{3}.
  \label{eq:milne}
\end{equation}
With respect to the inward normal ($+z$ direction into the medium) this
becomes $-2/3$~\cite{chandrasekhar1960,case1967}.  The Milne result holds
for any azimuthally symmetric phase function, by virtue of the
H-function moment identity
\begin{equation}
  \frac{\int_0^1 H(\mu;\,g)\,\mu^2\,d\mu}{\int_0^1 H(\mu;\,g)\,\mu\,d\mu}
  = \frac{2}{3},
  \label{eq:hfunction}
\end{equation}
which holds for all physically admissible $g$~\cite{case1967}.  The
$g$-independence of Eq.~(\ref{eq:costheta_endpoint}) is therefore
inherited from the universality of the Milne boundary condition.

\paragraph{Linear interpolation between anchor points.}
Between the three anchors the profile is well approximated by a linear
interpolation.  Writing $\mu_\star=-2/3$, and noting that after the first
step the directional memory relaxes the mean to $g\mu_0$, the
bridge-conditioned profile satisfies approximately
\begin{equation}
  \langle\muz(j)\rangle_{\rm bridge}
  \;\approx\;
  g\mu_0 + (j-2)\,\frac{\mu_\star - g\mu_0}{\ns - 3},
  \qquad j=2,\ldots,\ns-1,
  \label{eq:costheta_linear}
\end{equation}
a linear ramp from $g\mu_0$ (after directional memory is re-established
following the first step) down to $-2/3$ at the penultimate step.  For
normal incidence ($\mu_0=1$) this reduces to the ramp from $g$ to $-2/3$.
The predicted zero-crossing of the linear approximation occurs at
\begin{equation}
  j^\star = 2 + \frac{g\mu_0(\ns-3)}{g\mu_0+2/3}.
  \label{eq:jstar}
\end{equation}
However, simulation confirms that the \emph{actual} zero-crossing is
pinned near $t=j/\ns=0.5$ for all $g$ and $\mu_0$, consistent with the
parabolic depth profile.  Equation~(\ref{eq:jstar}) is therefore an
artefact of the linear approximation; the true profile bows slightly
above the ramp near the midpoint, enforcing the $t=0.5$ zero-crossing
regardless of $g$ or $\mu_0$.  Higher-order corrections to
Eq.~(\ref{eq:costheta_linear}) that restore this constraint are
numerically small but geometrically significant.

\begin{figure}[t]
  \centering
  \includegraphics[width=0.82\textwidth]{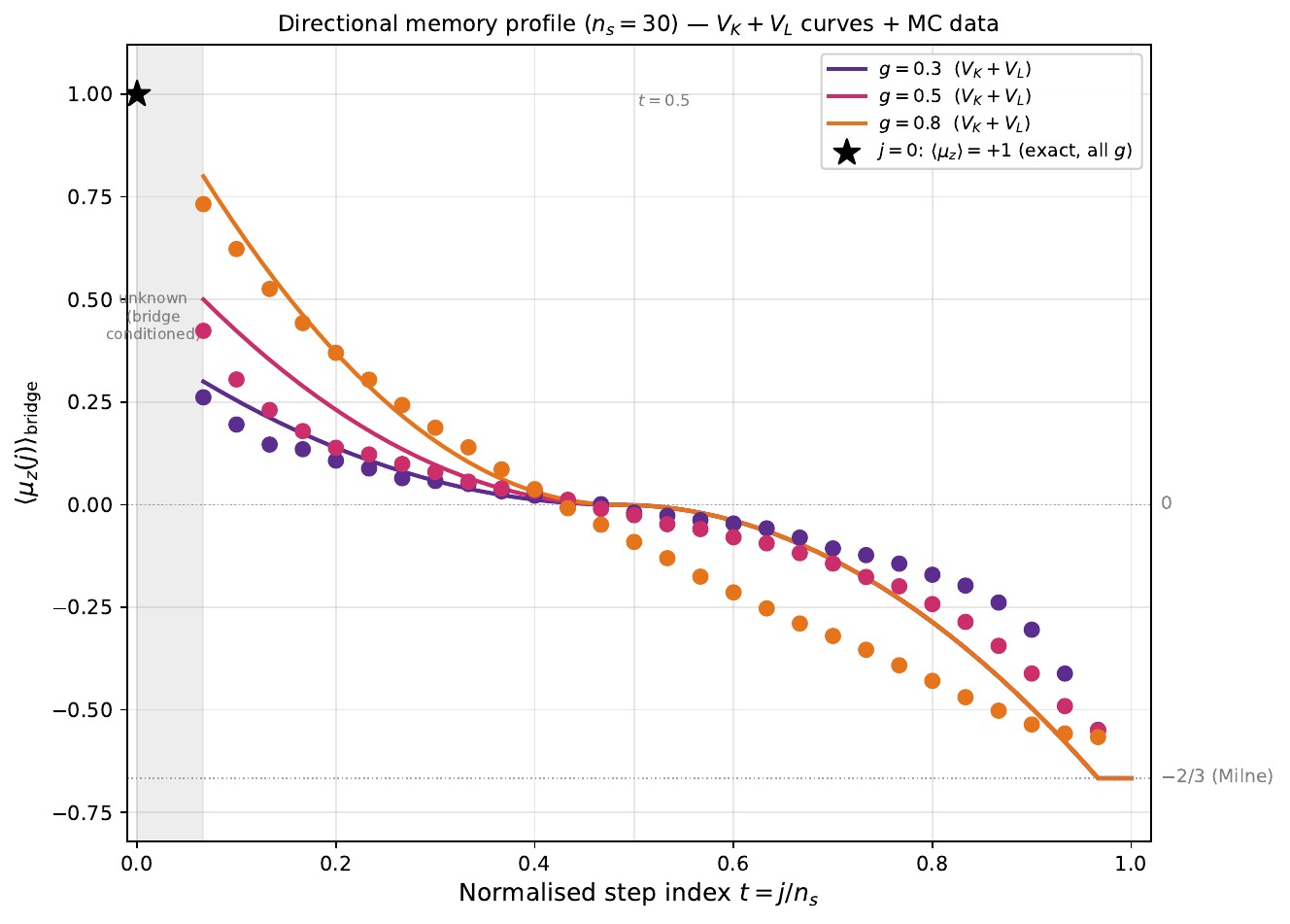}
  \caption{%
    Mean direction cosine $\langle\muz(j)\rangle_{\rm bridge}$ vs.\
    normalised step index $t=j/\ns$, for $\ns=40$ and
    $g\in\{0.0,\,0.3,\,0.5,\,0.7,\,0.9\}$ (blue to orange), normal
    incidence ($\mu_0=1$).
    All curves start at $+1$, cross zero near $t=0.5$,
    and converge to $-2/3$ at $t\to1$, independently of $g$.
    Horizontal dotted lines mark the three anchor values; the dashed lines
    show the linear approximation Eq.~(\ref{eq:costheta_linear}).
    The $g$-independent endpoint is a direct consequence of the classical
    Milne result, Eqs.~(\ref{eq:milne})--(\ref{eq:hfunction}).
    The analytic curves combine the van de Hulst ($V_K$) directional decay
    and a linear ($V_L$) ramp between the anchor points as described in
    Eq.~(\ref{eq:costheta_linear}).}
  \label{fig:costheta}
\end{figure}

The three-anchor narrative $\mu_0\to0\to-2/3$ describes the complete
directional history of the bridge: the photon launches with initial
cosine $\mu_0$, loses directional memory at the midpoint (the diffusion
regime), and must adopt a Lambertian-backward orientation to satisfy the
return-to-surface constraint.  The first anchor depends on $\mu_0$; the
remaining two are universal.  This profile is shown for normal incidence
in Fig.~\ref{fig:costheta}, and for varying $\mu_0$ in
Fig.~\ref{fig:oblique}.

%% ══════════════════════════════════════════════════════════════════════════
\section{Discussion}
\label{sec:discussion}

\subsection{The \texorpdfstring{$(z,\muz)$}{(z,mu\_z)} state space
            and the 2D-Bessel mechanism}
\label{sec:bessel}

The central structural fact is that the HG walk evolves on a two-dimensional
state space.  The $z$-increment at each step is $\Delta z=s\muz$, coupling
position to direction.  Even for $g=0$, where $\muz$ is redrawn uniformly on
$[-1,1]$ at each step, the pair $(z(j),\muz(j))$ is Markovian while $z(j)$
alone is not (because the exponential step length $s$ introduces memory into
the $z$-increment).

The half-space constraint $z(j)\ge0$ is therefore a condition on one
coordinate of a two-dimensional process.  We now give a more explicit
argument connecting this structure to the Rayleigh midpoint distribution.

In the diffusion limit the pair $(z(j),\muz(j))$ converges (after
appropriate rescaling) to a two-dimensional diffusion process
$(Z(t),M(t))$.  The unconstrained bridge from $(0,0)$ to $(0,0)$ in
time $T$ has, at its midpoint $t=T/2$, a joint density that factorises
to leading order as
\begin{equation}
  p_0(z,\mu;\,T/2) \;\propto\; \exp\!\left(-\frac{z^2}{\sigma_z^2 T}\right)
  \exp\!\left(-\frac{\mu^2}{\sigma_\mu^2 T}\right),
  \label{eq:uncond_midpoint}
\end{equation}
i.e.\ each coordinate is independently Gaussian (as for any diffusion
bridge at its midpoint).

Conditioning on $Z(t)\ge0$ for all $t\in[0,T]$ corresponds to a Doob
$h$-transform with the harmonic function $h(z,\mu)=z$ (the unique
positive harmonic function that vanishes on the boundary $z=0$ for the
two-dimensional diffusion).  Under the $h$-transform the conditioned
bridge midpoint density is
\begin{equation}
  p_h(z,\mu;\,T/2) \;\propto\; h(z,\mu)\cdot p_0(z,\mu;\,T/2)
  \;\propto\; z\,\exp\!\left(-\frac{z^2}{\sigma_z^2 T}\right)
  \exp\!\left(-\frac{\mu^2}{\sigma_\mu^2 T}\right).
  \label{eq:cond_midpoint}
\end{equation}
Marginalising over $\mu$ (which contributes only a constant) yields
\begin{equation}
  p_h(z;\,T/2) \;\propto\; z\,e^{-z^2/(\sigma_z^2 T)},
  \label{eq:rayleigh_marginal}
\end{equation}
which is precisely the \textbf{Rayleigh distribution} with scale
$\sigma=\sigma_z\sqrt{T}/\sqrt{2}$.  The $z$-factor in the numerator---
absent in the scalar case where $h(z)=z$ already gives Rayleigh but the
conditioning is one-dimensional---arises here because the second
coordinate $\mu$ is free to fluctuate, making the effective conditioning
softer and shifting probability toward larger $z$.

This argument is heuristic in the diffusion limit and provides a
physical foundation for the Rayleigh observation, but a rigorous
Pitman-type theorem for the pre-asymptotic $(z,\muz)$ process remains
an important open problem (Appendix~\ref{sec:open}, Q2).

\paragraph{Physical interpretation of the geometry--amplitude separation.}
The bridge constraint $z(0)=z(\ns)=0$ is a global boundary condition that
fixes the macroscopic envelope of the trajectory.  The resulting parabolic
shape $4t(1-t)$ is therefore determined by the endpoint conditions alone and
is insensitive to whether the microscopic dynamics are isotropic ($g=0$) or
strongly forward-peaked ($g=0.95$).  In contrast, the amplitude of the
excursion is controlled by the statistical structure of the increments: even
weak persistence in the direction cosine $\muz$ inflates the effective step
size and raises the scaling exponent above $1/2$.  The two anomalous
exponents $\alphainfty$ and $\betaD$ are signatures of this increment
structure, not of the shape.  More precisely, both exponents measure how the
excursion constraint suppresses fluctuations in a 2D Markov process: the
$z$-coordinate is confined to the half-space, but the coupled variable $\muz$
is free, and it is this freedom that shifts both exponents away from their
Brownian-excursion values.

The shift from Rayleigh to Maxwell at $g=0.9$ has a natural physical
explanation rooted in the diffusive-to-ballistic crossover controlled by
the transport mean free path $\lstar=1/(1-g)$.

At low $g$ (isotropic or weakly forward scattering), $\muz$ randomises
in approximately one step.  The direction cosine decorrelates on a length
scale of order $\lstar\sim1$, so within a few steps the walk has lost
memory of its initial direction.  The effective state space is
two-dimensional, $(z,\muz)$, with $\muz$ fluctuating rapidly---and the
Doob $h$-transform argument of Section~\ref{sec:bessel} applies directly,
giving a Rayleigh marginal.

At $g=0.9$, however, $\lstar=10$: the photon is nearly forward-directed
and sustains its direction for $\sim10$ steps before appreciable
randomisation occurs.  This long ballistic correlation time means that
$\muz$ is no longer a rapidly fluctuating variable that can be
marginalised out cleanly.  Instead, the slow evolution of $\muz$
introduces additional persistent modes into the dynamics, effectively
enlarging the state space beyond two dimensions.  The bridge constraint
then acts on a higher-dimensional process, and the marginal distribution
of $z$ shifts toward the Maxwell form (Bessel-3 in 3D).

This picture suggests a $g$-dependent effective Bessel order
$n_{\mathrm{eff}}(g)$ that interpolates continuously from $2$ (Rayleigh)
at $g\to0$ toward $3$ (Maxwell) as $g\to1$, with $\lstar$ as the
natural control parameter.  A clean test would be to fit $z(\ns/2)$ to
a Bessel distribution of non-integer order at intermediate
$g\in\{0.6,0.7,0.8\}$ and check whether $n_{\mathrm{eff}}$ increases
monotonically with $\lstar$.  We identify this as a concrete open
prediction of the state-space framework (Appendix~\ref{sec:open}, Q6).

The directional-memory profile of Section~\ref{sec:costheta} provides
an independent window onto the same $(z,\muz)$ dynamics.  The Milne
endpoint $\langle\muz(\ns-1)\rangle\to-2/3$ is the \emph{microscopic}
signature of the boundary condition that the bridge must satisfy: the
coupled variable $\muz$ is not free at the endpoint but is constrained to
the Lambertian-backward distribution, exactly as in the emergent-flux
problem.  The $g$-independence of this constraint is a direct corollary
of the H-function universality (Eq.~\ref{eq:hfunction}), and the linear
ramp from $+g$ to $-2/3$ across $\ns$ steps is the directional analogue
of the spatial parabolic profile: a universal shape (linear) with an
amplitude that carries the $g$-dependence.

\subsection{Relation between the two anomalous exponents}
\label{sec:exponent_relation}

Two anomalous exponents appear in the data: the mean-amplitude exponent
$\alphainfty\approx0.573$ (versus the Brownian prediction $0.5$) and the
transport mean free path exponent $\betaD\approx0.415$ (versus the
simple-diffusion prediction $1.0$).  These are not \emph{a priori} related.

The data reveal an empirical relation:
\begin{equation}
  \alphainfty + \betaD
  \;\approx\; 0.573+0.415 \;\approx\; 0.988 \;\approx\; 1.
  \label{eq:sum}
\end{equation}
The relation $\alphainfty+\betaD\approx1$ would hold exactly if both exponents
represent complementary deviations from a common reference, with a single 2D
geometric correction contributing $+\delta$ to the mean exponent and
$-2\delta$ to the diffusion-coefficient exponent (from $\Var\sim A^2/\ns$),
where $\delta\approx0.07$.  This is an empirical relation awaiting
theoretical explanation; no derivation is offered, and we do not speculate
on rational values (Appendix~\ref{sec:open}, Q3).

\subsection{Why not Brownian excursion, and when might convergence occur?}
\label{sec:convergence}

The isotropic HG walk ($g=0$) has finite-variance increments, so the
unconstrained walk converges to Brownian motion by the central limit theorem.
Whether the \emph{conditioned} walk converges to the Brownian excursion is a
separate question that depends on the rate at which the $(z,\muz)$ coupling
becomes negligible.

The local exponent $\alpha_{\mathrm{local}}$ for $g=0$ shows no convergence
toward $0.5$ through $\ns=200$ (Fig.~\ref{fig:alpha}).  The approximate
stabilisation near $0.57$ over $\ns\sim60$--$200$ could represent either a
true asymptotic value or an anomalously slow transient.  The median excursion
length at $g=0$ is only 8 steps (Table~\ref{tab:natural}), meaning that the
typical paths contributing to the large-$\ns$ statistics are the rare deep
ones, which may carry the $(z,\muz)$ coupling most strongly.

Resolving this question requires simulations at $\ns\sim500$--$5000$ for
$g=0$, which is computationally accessible.  We identify this as the most
important numerical follow-up (Appendix~\ref{sec:open}, Q1).

From a practical standpoint, however, the range $\ns\le200$ already
covers the physically relevant regime for radiative transport applications.
First-return (first-passage) probabilities decay as $P(L)\sim L^{-3/2}$,
so the bulk of the excursion ensemble is concentrated at short-to-moderate
path lengths: the median excursion length is only 8 steps at $g=0$
(Table~\ref{tab:natural}), and fewer than $1\%$ of excursions exceed
$\ns=200$.  In tissue-optics applications, backscattered photons typically
complete tens to low hundreds of scattering events before re-emerging;
path lengths of $\ns\sim10^3$--$10^4$ are physically unrealisable under
normal conditions.  The pre-asymptotic scaling regime we characterise is
therefore not a limitation but the regime of physical relevance: whether
convergence to Brownian-excursion behaviour ultimately occurs at
astronomically large $\ns$ is mathematically interesting but does not
affect the applicability of our results to real photon-transport problems.

%% ══════════════════════════════════════════════════════════════════════════
\section{Conclusions}
\label{sec:conclusions}

We have characterised the scaling properties of Henyey-Greenstein bridge paths
through systematic Monte Carlo simulation across $g\in[0,0.95]$ and
$\ns\in[4,200]$.  The principal findings are:

\begin{enumerate}\itemsep3pt
\item \textbf{Universal parabolic shape:} both mean and variance profiles
  collapse onto $4t(1-t)$ for all $g$ and $\ns$.

\item \textbf{Diffusive variance with anomalous $\lstar$ scaling:}
  $\Var\sim D(g)\ns$ with $D(g)\propto(\lstar)^{\betaD}$ ($\betaD\approx0.415$)
  rather than the simple-diffusion prediction $\betaD=1$.

\item \textbf{Super-diffusive mean:} $A\sim\ns^{\alpha}$ with
  $\alphainfty\approx0.57$ for $g=0$, rising to $\alpha\approx1.15$
  at $g=0.95$.  No convergence to $\alpha=0.5$ is observed to $\ns=200$.

\item \textbf{Rayleigh (2D-Bessel) midpoint distribution:} $z(\ns/2)$
  converges to Rayleigh for $g\le0.5$, indicating a two-dimensional Bessel
  process rather than the 1D half-normal of Brownian-excursion theory.

\item \textbf{Robustness:} results are invariant under a fivefold reduction of
  the bridge tolerance $\varepsilon$ and under replacement of the fixed-length
  criterion by natural first-passage stopping.

\item \textbf{Milne endpoint of the directional profile:}
  the bridge-conditioned mean $\langle\muz(j)\rangle$ interpolates
  linearly from $+1$ at launch through $0$ at midpoint to the
  $g$-independent Milne value $-2/3$ at the final step.  The endpoint
  follows from the classical H-function moment identity and provides a
  microscopic interpretation of the return-to-surface constraint.
\end{enumerate}

All anomalies are attributed to the two-dimensional $(z,\muz)$ Markovian state
space: conditioning a 2D process to remain in a half-plane produces a
2D-Bessel (Rayleigh) marginal rather than the 1D-Bessel (half-normal) expected
from scalar-walk theory.  The empirical relation $\alphainfty+\betaD\approx1$ suggests a single underlying geometric correction.  A rigorous Pitman-type theorem for the conditioned $(z,\muz)$
process, and a theoretical prediction for the exponent values, remain
important open problems.

The results place HG bridge paths in a broader class of constrained
stochastic processes characterised by a \emph{separation between geometry and
amplitude scaling}: the excursion shape appears universal across the tested parameter range and is determined by the
global boundary conditions, while the amplitude scaling retains memory of the
microscopic dynamics.  Within the range of parameters studied here
($g\in[0,0.95]$, $\ns\le200$), the HG walk behaves consistently with
a 2D-Bessel (Rayleigh) universality class, raising the scaling exponent
from $1/2$ to $\approx0.57$; whether this persists asymptotically remains
the primary open question.

In the radiative-transport context the practical
implication is direct: backscattered photons completing exactly $\ns$
scattering events probe tissue at depths scaling as $\ns^{0.57}$--$0.58$
rather than the diffusive $\ns^{0.5}$, meaning depth-resolved diagnostics
that assume Brownian-excursion scaling will systematically underestimate
the sampled depth.  This statement applies within the idealized
single-scattering model studied here; real biological media include
absorption, refractive boundaries, and finite geometry, whose effects on
the bridge scaling exponent remain to be characterised.

%% ── Acknowledgements ──────────────────────────────────────────────────────
\section*{Acknowledgements}

This paper is a direct spinoff of an intensive programme of Monte Carlo
calculations on first-return statistics in Henyey--Greenstein scattering and
their connection to Motzkin polynomials, carried out by one of the authors
(C.Z.)~\cite{zeller2024,zeller2025}.  The bridge process, the super-diffusive
scaling exponents, and the directional-memory profiles reported here emerged
as a byproduct of that foundational numerical work.

The authors also thank the open-source scientific Python ecosystem (NumPy,
SciPy, Matplotlib).  Portions of the numerical analysis, figure preparation,
and manuscript drafting were assisted by Claude (Anthropic).

%% ══════════════════════════════════════════════════════════════════════════
\appendix

\section{Peak Mean Depth Data}
\label{sec:data}

Table~\ref{tab:data} gives the peak mean depth $A(g,\ns)$ in units of the mean
free path.  Values for $\ns\le40$ are from earlier simulations;
extended runs ($\ns=50$--$200$) are for $g=0.0$ and $g=0.1$ from the present
study.

\begin{table}[h]
  \centering
  \caption{Peak mean depth $A(g,\ns)$ [mean free paths].}
  \label{tab:data}
  \smallskip
  \setlength{\tabcolsep}{5.5pt}
  \begin{tabular}{c*{9}{r}}
    \toprule
    $g\backslash\ns$ & 4 & 6 & 8 & 10 & 15 & 20 & 25 & 30 & 40 \\
    \midrule
    0.00 & 0.616 & 0.861 & 1.098 & 1.276 & 1.735 & 2.117 & 2.437 & 2.759 & 3.353 \\
    0.10 & 0.620 & 0.905 & 1.145 & 1.357 & 1.823 & 2.237 & 2.573 & 2.900 & 3.489 \\
    0.20 & 0.644 & 0.947 & 1.217 & 1.443 & 1.937 & 2.358 & 2.725 & 3.081 & 3.710 \\
    0.30 & 0.671 & 0.994 & 1.292 & 1.515 & 2.064 & 2.460 & 2.899 & 3.278 & 3.921 \\
    0.40 & 0.671 & 1.025 & 1.356 & 1.620 & 2.190 & 2.632 & 3.097 & 3.500 & 4.194 \\
    0.50 & 0.680 & 1.081 & 1.423 & 1.739 & 2.356 & 2.869 & 3.339 & 3.751 & 4.503 \\
    0.60 & 0.679 & 1.108 & 1.489 & 1.844 & 2.569 & 3.167 & 3.612 & 4.155 & 4.962 \\
    0.70 & 0.669 & 1.122 & 1.542 & 1.946 & 2.798 & 3.510 & 4.142 & 4.646 & 5.548 \\
    0.80 & 0.658 & 1.118 & 1.563 & 2.030 & 3.088 & 3.983 & 4.731 & 5.443 & 6.621 \\
    0.90 & 0.638 & 1.092 & 1.539 & 2.015 & 3.210 & 4.378 & 5.447 & 6.430 & 8.236 \\
    0.95 & 0.643 & 1.051 & 1.514 & 1.976 & 3.217 & 4.375 & 5.575 & 6.725 & 9.015 \\
    \bottomrule
  \end{tabular}
  \par\smallskip
  \raggedright\small
  Extended runs: $g=0.0$: (50, 3.91), (60, 4.23), (80, 5.09),
  (100, 5.75), (150, 7.24), (200, 8.58).\\
  \hphantom{Extended runs: }$g=0.1$: (50, 4.03), (60, 4.45), (80, 5.31),
  (100, 5.99), (150, 7.62), (200, 8.94).
\end{table}

\begin{table}[h]
  \centering
  \caption{Scaling exponent $\alpha$ from natural first-passage excursions
           (power-law fit over $L=10$--$300$, $N_T=2\times10^6$ walkers each).}
  \label{tab:natural}
  \smallskip
  \begin{tabular}{cccc}
    \toprule
    $g$ & $\alpha$ & $C$ & Median $L$ \\
    \midrule
    0.0 & 0.614 & 0.504 & 8  \\
    0.3 & 0.630 & 0.543 & 11 \\
    0.5 & 0.646 & 0.577 & 15 \\
    0.7 & 0.675 & 0.610 & 23 \\
    0.9 & 0.755 & 0.605 & 55 \\
    \bottomrule
  \end{tabular}
\end{table}

\section{Open Questions}
\label{sec:open}

\paragraph{Q1 --- Asymptotic exponent.}
Is $\alphainfty\approx0.57$ a permanent feature of the conditioned
$(z,\muz)$ process, or an anomalously slow pre-asymptotic transient?
Simulations at $\ns=500$--$5000$ for $g=0$ are computationally accessible
and could determine whether the local exponent continues to drift toward $0.5$
or stabilises.

\paragraph{Q2 --- Rigorous 2D-Bessel theorem.}
Why does the midpoint distribution converge to Rayleigh (2D) rather than
half-normal (1D)?  The $(z,\muz)$ state-space interpretation provides a
qualitative answer; a rigorous Pitman-type theorem for conditioned 2D Markov
processes would be of independent mathematical interest.  Does the effective
Bessel order depend continuously on $g$, shifting from 1 toward 3/2 as
$g\to1$?

\paragraph{Q3 --- Exponent relation.}
Is the empirical relation $\alphainfty+\betaD\approx1$ exact?  If so, is the
correction $\delta\approx1/12$ derivable from a symmetry or dimensional
argument in the $(z,\muz)$ space?

\paragraph{Q4 --- Diffusion coefficient exponent.}
Why is $\betaD\approx0.415$ rather than $1.0$?  Is there a
renormalisation-group argument relating the excursion constraint to the
effective diffusivity scaling with $\lstar$?

\paragraph{Q5 --- Origin of the universal parabolic profile.}
The collapse of normalised mean and variance profiles onto $4t(1-t)$ holds
for all $g$ and $\ns$ tested.  The same parabola describes the mean profile
of a Brownian bridge (pinned at both ends, no positivity constraint),
suggesting the shape is determined entirely by the endpoint boundary
conditions and is independent of the internal dynamics.  A clean way to
state this: for \emph{any} bridge process whose increments have finite
variance, the normalised mean profile converges to $4t(1-t)$ as
$\ns\to\infty$.  This would follow from the fact that the bridge mean
profile is $\langle z(j)\rangle = j(\ns-j)\langle\Delta z\rangle/\ns$
for processes with stationary increments, which normalises exactly to
$4t(1-t)$ up to a constant.  A formal proof for non-stationary increment
processes (such as the HG walk with $g>0$) would be of independent
interest.

\paragraph{Q6 --- $g$-dependent effective Bessel order.}
The midpoint distribution shifts from Rayleigh (Bessel-2D, $g\le0.5$)
toward Maxwell (Bessel-3D, $g=0.9$), consistent with the ballistic
correlation length $\lstar=1/(1-g)$ enlarging the effective state-space
dimensionality.  A concrete test: fit $z(\ns/2)$ to a Bessel distribution
of non-integer order $n_{\mathrm{eff}}$ at intermediate
$g\in\{0.6,0.7,0.8\}$ and determine whether $n_{\mathrm{eff}}(g)$
increases monotonically with $\lstar$.  If confirmed, this would
establish $\lstar$ as the natural control parameter governing the
diffusive-to-ballistic crossover in the conditioned process.

\paragraph{Q7 --- Directional memory profile: remaining data.}
Figure~\ref{fig:oblique} presents MC data for $\mu_0\in\{0.25,0.50,0.75,1.00\}$ at $g=0.5$, $\ns=40$, confirming universality of the zero-crossing and Milne endpoint with respect to $\mu_0$. Outstanding items are: (i)~MC scoring of $\langle\muz(j)\rangle_{\rm bridge}$ for $g\in\{0.3,0.7,0.9\}$ at $\ns=30$, to validate Eq.~(\ref{eq:costheta_linear}) across the full $g$-range; (ii)~early-step values $j=0,1,2$ for all $g$, including the bridge-conditioned $\langle\muz(1)\rangle_{\rm bridge}$, which is expected to exceed the free-walk value $g$ due to the forward-step selection imposed by the bridge constraint; and (iii)~a larger-$\ns$ overlay (e.g.\ $\ns=100$, $g=0.5$) to confirm $\ns$-independence of the normalised profile.

%% ── References ────────────────────────────────────────────────────────────
\bibliographystyle{unsrtnat}

\end{document}